\documentstyle[12pt,psfig,epsf]{article}
\newcommand{\be}{\begin{equation}}
\newcommand{\ee}{\end{equation}}

\begin{document}
\hfill  NRCPS-HE-2000-13

\vspace{24pt}
%\begin{titlepage}
%\title{
\begin{center}
{\large \bf The system with exponentially degenerate vacuum state

}%title ends

\vspace{24pt}
%\author{ 
{\sl  George K. Savvidy}\\
National Research Center Demokritos, \\
Agia Paraskevi, GR-15310  Athens, Greece \\
{\tt email:savvidy@argo.nuclear.demokritos.gr}
\vspace{1cm}

%}%author ends
%}
%\date{}%in order NOT to write the date
%\maketitle
\end{center}
\vspace{60pt}

\centerline{{\bf Abstract}}

\vspace{12pt}
\noindent

I suggest and examine artificial material which has exponentially 
degenerate vacuum state. The corresponding 
Hamiltonian  contains only exotic four-spin interaction term \cite{sav2}. 
Each vacuum state is realized as a particular spin configuration  
separated from others by potential barriers. 
The benefit of such system in practical applications is 
that it can be used as high density magnetic recording system 
which can reduce storage of one bit information to $nm$ scale. 
The information is stored as a particular vacuum state of the system.  
The process of recording can be visualized 
as a process in which the system  moves from one vacuum state to another.
Storing information in the form of different  vacuum states 
separated by potential barriers will allow to protect it 
from fluctuations and for a longer time. 
These  materials can be realized as 
lattices of nuclear spins with specially adjusted interactions.
The planes of flipped spins  can in principle be of atomic scale.

\bigskip
\vskip 60 true pt
\centerline{\it 
Dedicated to the memory of Professor Judah Eisenberg }

%\end{abstract}
%\thispagestyle{empty}
%\end{titlepage}

\newpage

\pagestyle{plain}
%\pagenumbering{roman}
\section{Introduction}

%\vspace{.5cm}

Lattice spin systems and different ferromagnetic materials with 
competing interaction have been catching attention in the last decades 
\cite{domb}. Spin glasses, alloys and amorphous systems which have
randomly distributed competing interaction, are also of similar 
nature \cite{fischer,bohn,sherrington}. 

In the recent articles \cite{weg,sav1,sav2,weg1} 
the authors formulated a new class of statistical systems 
in which the energy functional is proportional to the total 
length of edges of the interface \cite{amb}.
These lattice spin systems have 
specially adjusted interaction between spins in order to simulate 
a given energy functional. The specific property of these
systems is that they have very high -- exponential degeneracy of the ground state.
This happen simple because surface tension forces are tuned to vanish \cite{amb}.
This peculiar property of the system could make them useful for practical 
applications. In this article I suggest and examine application of  this
system in memory devices and possibly to store bits in future quantum computers. 

\begin{figure}
\centerline{\hbox{\psfig{figure=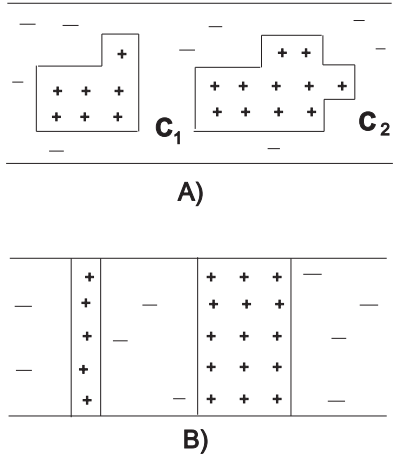,height=6cm,angle=00}}}
\caption[fig1]{ Magnetic strip of the width $h$ with islands $C_{i}$ of flipped spins.
For the 3D-Ising ferromagnet the energy of this configuration is 
equal to the length of the boundary $C_{i}$ times the strip width $h$ and is 
proportional to the area $S=h \sum_{i} length(C_{i})$ of the interface. 
It is nonzero for both configurations A and B.
For the gonihedric system (\ref{hamil}) the energy is proportional to the 
curvature, of the boundary and of the intersections, multiplied by the strip width $h$.
This is the size of the interface \cite{amb}
$A=h [\sum_{i} (Right~Angles)_i+  4  \kappa \sum_{i}~(Intersections)_i]$. 
It is nonzero for the configuration A,  but is equal to zero for the 
configuration B,  the configuration B is one of the ground states.}
\label{fig1}
\end{figure}

\begin{figure}
\centerline{\hbox{\psfig{figure=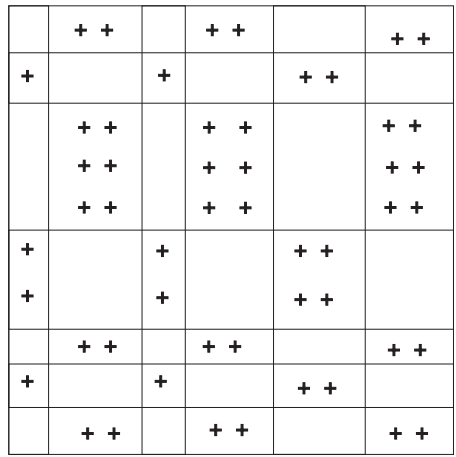,height=6cm,angle=-90}}}
\caption[fig2]{ Magnetic strip of the width $h$ with islands $C_{i}$ of flipped spins.
For the gonihedric system with $k=0$ the energy is proportional to the curvature
of the boundary times $h$, $A=h \sum_{i} (Right~Angles)_i$, but  now without 
contributions coming from 
self-intersections, therefore any "chess-board" configuration is a ground state.}
\label{fig2}
\end{figure}

In three dimensions the corresponding Hamiltonian is equal to 
\cite{sav1}
\be
H_{gonihedric}^{3d}=- 2k \sum_{\vec{r},\vec{\alpha}} \sigma_{\vec{r}}
\sigma_{\vec{r}+\vec{\alpha}}
+ \frac{k}{2} \sum_{\vec{r},\vec{\alpha},\vec{\beta}} \sigma_{\vec{r}}
\sigma_{\vec{r}+\vec{\alpha} +\vec{\beta}}
-  \frac{1-k}{2} \sum_{\vec{r},\vec{\alpha},\vec{\beta}} \sigma_{\vec{r}}
\sigma_{\vec{r}+\vec{\alpha}} \sigma_{\vec{r}+\vec{\alpha}+\vec{\beta}}
\sigma_{\vec{r}+\vec{\beta}}, \label{hamil}
\ee
where $\vec r$ is a three-dimensional vector on the lattice 
$Z^{3}$, the  components of which are integer and $\vec \alpha$, $\vec 
\beta$ are unit vectors parallel to the axes. 
This lattice system crucially depends on the coupling constant $k$ 
called self-intersection coupling constant.
The form of the Hamiltonian $H^{k}$
and the symmetry of the system essentially depends on k: 
when $k \neq 0$  one can flip spins on arbitrary parallel layers 
and thus the degeneracy of the ground state is equal to $3\cdot 2^{N}$, 
where $N^3$ is the size of the lattice (see Figures \ref{fig1}). 
When $k=0$ the system
has even higher symmetry, {\it all states, including the ground
state are exponentially degenerate} \cite{sav2} 
\footnote{This is very important property of the system 
which allows also to construct the dual system \cite{sav2}. }. 
This degeneracy is equal to $2^{3N}$ \cite{sav2}. This is  because now 
one can flip  spins on arbitrary layers, even on
intersecting ones (see Figure \ref{fig2}). The corresponding 
Hamiltonian  contains only exotic four-spin interaction term. 

This simply means that  the "crystal" of the size $N^{3}$ has 
$2^{3N}$ different ground states.  This  
exponential degeneracy is much bigger than the degeneracy of the
vacuum state of the Ising ferromagnet, which is simply equal to two,  
and in this respect has very close nature with spin glasses
\cite{bathas} and may describe liquid-glassy phase transition \cite{johnson,felix}. 

In the usual Ising ferromagnet we have two different vacuum states, 
so in order to store more than one bit of information one should allow  
excited states as it is shown  on Figure \ref{fig1} A.  
With decreasing geometries of recording and reading heads 
and increasing magnetic media storage densities, these excitations 
become metastable thanks to the fluctuations on $nm$ scale 
and thus can not be protected from damages. 
In about 10 years, storage of one bit of information 
is expected to cover an area of 100x100 $nm^2$, and metastability
of excitations on $nm$ scale becomes an increasingly limiting 
factor of performance \cite{nano}.

Opposite to that situation the "crystal" of  size $N^3$  
which has special interaction between spins 
can "memorize"  $2^{3N}$ different 
states, which are well separated by potential barriers 
(see Figure \ref{fig1} and \ref{fig2}). Therefore we suggest that
natural or artificial materials with  the corresponding structure 
of interactions can be used as magnetic recording systems.  
This article considers  the question of possible construction of
artificial material with the above property, its possible 
realization and application in memory devices and 
possibly to store bits in future quantum computers.

\section{System with $3 \times 2^N$ ground state degeneracy}

The benefit of having system with exponentially degenerate vacuum 
state in practical applications is 
that it can be used as high density magnetic recording system. 
Each vacuum state is realized as a particular spin configuration  
separated from others by potential barriers of height $U$ 
which is proportional 
to the width of the magnetic strip $h$ (see formulas (\ref{height}) and 
Figure \ref{fig1}). 
The information can be stored as a particular vacuum state of the system. 
The process of recording can be visualized 
as a process in which the system  moves form one vacuum state to another.
Storing information in the form of different  vacuum states 
separated by potential barriers will allow to keep it  
safely away from fluctuations and for a longer time. 
We shall demonstrate that this  material can be realized as a
lattice of nuclear spins with specially adjusted interactions.

First let us consider the system which has exponentially degenerate vacuum 
state, but only at zero temperature. 
This "crystal" which has $3 \times 2^N$ ground state degeneracy has been
constructed in \cite{weg} and was studied in a number of articles 
analytically \cite{sav1,sav2,weg1,thordur} and numerically \cite{bathas,johnson}
(see Figure \ref{fig1}).
It corresponds to the case $k=1$  in the equation (\ref{hamil}). 
The Hamiltonian of the system has the form \cite{weg}

\be
H_{gonihedric}^{3d}=- J \sum_{\vec{r},\vec{\alpha}} \sigma_{\vec{r}}
\sigma_{\vec{r}+\vec{\alpha}}
+ \frac{J}{4} \sum_{\vec{r},\vec{\alpha},\vec{\beta}}
\sigma_{\vec{r}} \sigma_{\vec{r}+\vec{\alpha}+\vec{\beta}}.     \label{k1}
\ee
The energy of a configuration is equal to 
the curvature of the boundary plus the energy at the intersections
\be
E=h [\sum_{i} (Right~Angles)_i  + 4 \kappa \sum_{j} (Intersections)_j]. \label{height}
\ee
In this case the Hamiltonian includes only competing ferromagnetic 
and antiferromagnetic interactions. The ferromagnetic coupling constant
is $J_{ferromagnetic} = J$ and the antiferromagnetic coupling constant 
should be four times smaller $J_{antiferromagnetic} = J/4$, 
thus the ratio is equal to four
\be
J_{ferromagnetic}/J_{antiferromagnetic} = 4. 
\ee
The critical temperature $\beta_c =J/KT \approx 0.44$ has 
been predicted in \cite{sav2} (see also \cite{thordur})
and confirmed by  Monte-Carlo simulations
\cite{bathas,johnson}  and the low temperature 
expansion \cite{weg1}, thus $T_c \approx  J/ 0.44K$. 
In order to have the phase transition point at high temperatures 
the coupling constant $J$ should be large  enough. 

It is an interesting question if there exist a material with the 
above interactions. The crystalline $EuS$ and $Eu_{x}Sr_{1-x}S$ 
\cite{bohn}, which is a ferromagnetic insulator, has exchange energy 
coupling constants equal to
$$
J_{ferromagnetic} = (0.221 \pm 0.003) K,
$$
$$
J_{antiferromagnetic} = ( 0.1 \pm 0.004) K,
$$
therefore the ratio is equal to two and one should look for 
other materials with appropriate coupling constants.  
If the material with these interactions will be found or
constructed it will be not so useful for direct applications because the 
exponential degeneracy of the vacuum state  is
lifted at nonzero temperatures. This is because 
nonzero surface tension is generated
by quantum-thermal  fluctuations \cite{amb,johnson,weg1}, 
the area term in the energy functional.
This effect will suppress the interface walls and thus the degeneracy
is lifted. 
From other side, crystal of this type can be helpful 
for experimental verification of the string tension generation 
phenomena in string theory suggested in \cite{amb}, sort of 
experimental laboratory for QCD string.

\section{System with $ 2^{3N}$ ground state degeneracy}

The system which has even higher degeneracy of the ground state than the one 
which we described in the previous section has been constructed in
\cite{sav2}. The advantage of this system is that the degeneracy of the vacuum 
state remain untouched even at nonzero temperatures. 

In terms of Ising spin variables $\sigma_{\vec r}$ the Hamiltonian of 
the system with $ 2^{3N}$ ground state degeneracy can be written
in the form  \cite{sav2}
\be
H^{3D}_{Gonihedric} = -J_{4} \sum_{\vec r,\vec \alpha,\vec \beta}
\sigma_{\vec r} \sigma_{\vec r+\vec \alpha}
\sigma_{\vec r+\vec \alpha+\vec \beta}
\sigma_{\vec r+\vec \beta} \label{fourspin}
\ee
where $\vec r$ is a three-dimensional vector on the lattice 
$Z^{3}$, the  components of which are integer 
and $\vec \alpha$, $\vec \beta$ are unit vectors parallel to the axes. This
Hamiltonian corresponds to the case $k=0$ in equation (\ref{hamil}).
We should stress that the Ising spins in (\ref{fourspin}) are on the vertices 
of the lattice $Z^{3}$ and are not on the links and the 
coupling constant $J_{4}$ should be $positive$. The
Hamiltonian  contains only exotic four-spin interaction term 
$\sigma \sigma\sigma\sigma$. 

It is hard to see how this four-spin interaction term can be 
simulated even by artificial materials. In this section we propose 
to introduce additional spin which is located in the center of 
plaquette and then  adjust it interaction so that effective 
interaction between four-spins located at the vertices of the 
plaquette will be of the form $\sigma \sigma\sigma\sigma$. One can
consider this transformation as a modification of the {\it decoration
transformation} and it is analogous to the {\it star-triangle transformation}
\cite{wannier}.

\begin{figure}
\centerline{\hbox{\psfig{figure=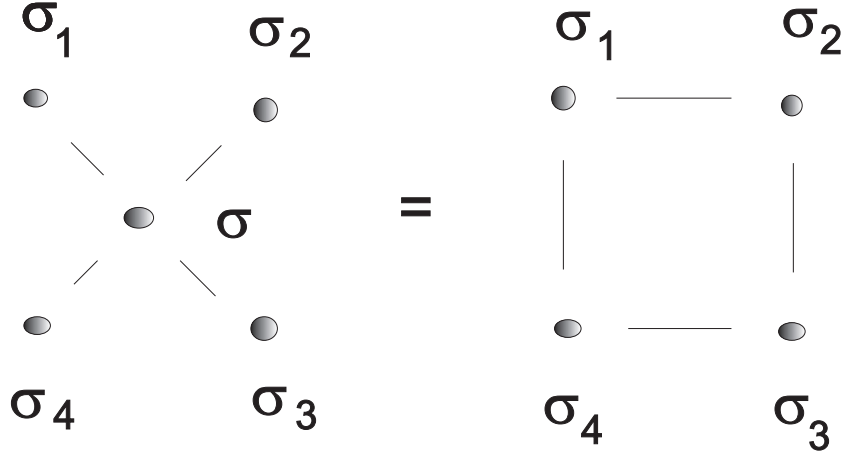,height=6cm,angle=00}}}
\caption[fig3]{Intergation over the central spin $\sigma$ produces 
an effective interaction between four spins at the vertices of the 
plaquette.}
\label{fig3}
\end{figure}

Thus to generate four-spin interaction term we shall introduce
central spin $\sigma$ which interacts with its four neighbors 
(see Figure \ref{fig3}). We should prove that integrating out 
the interaction with 
the central spin will generate the necessary four-spin interaction term.
Therefore we have to prove the existence of the following relation
\be
\frac{1}{2} \sum_{\sigma = \pm 1} e^{J \sigma (\sigma_{1}+ \sigma_{2}+ 
\sigma_{3}+ \sigma_{4} )} = 
A e^{J_{1} ( \sigma_{1} \sigma_{2} + 
\sigma_{2}\sigma_{3}+\sigma_{3}\sigma_{4} + \sigma_{4}\sigma_{1})}
e^{J_{2} (\sigma_{1}\sigma_{3} + \sigma_{2}\sigma_{4})}   
e^{J_{4} (\sigma_{1}\sigma_{2}\sigma_{3}\sigma_{4})} \label{dualtran}
\ee
with nonzero coupling constant $J_{4}$. Here $\sigma$ is the central 
spin and $\sigma_{i}, i=1,2,3,4$ are spins on the vertices. If this
relation holds then it means that the interaction of the  central spin
with its neighbors can be effectively replaced by direct $J_{1}$, 
diagonal$J_{2}$ and four-spin interaction  $J_{4}$ (see Figure \ref{fig3}). 
To express coupling constants $J_{1},J_{2},J_{4}$ and $A$ through the 
coupling constant $J$ we have to solve the system of $2^{4}$ 
equations which appear when we substitute the values of the spins
$\sigma_{i}, i=1,2,3,4$ into the equation (\ref{dualtran}). Only four
equations are independent because of the global $Z_{2}$ invariance:
\begin{eqnarray}
ch(4J) = A exp(4J_{1} +2 J_{2} + J_{4} ),    \nonumber\\
1= A exp(-2 J_{2} +  J_{4}),~~~~~~~~~\nonumber\\
ch(2J) = A exp(- J_{4} ),~~~~~~~~~~\nonumber\\
1=  A exp(-4J_{1} +2 J_{2} J_{4} ),~~~~~~~~\label{con1}
\end{eqnarray}
From the first,second and the forth equations it follows that
$$
J_{1} = J_{2}
$$
and our equations reduce to:  
\begin{eqnarray}
ch(4J) = A exp(6 J_{2} + J_{4} ),\nonumber\\
1= A exp(-2 J_{2} +  J_{4}),~~~\nonumber\\
ch(2J) = A exp(- J_{4} ).~~~ \label{con2}
\end{eqnarray}
From these equations it follows that:
$$
ch(4J)= A^{8}/(ch 2J)^4
$$
and thus
\be
A = (ch^{4}(2J)*ch(4J))^{1/8}.  \label{A}
\ee
Using again equations (\ref{con2}) we can find coupling 
constants $J_2$ and $J_4$
\begin{eqnarray}
J_{2}= \frac{1}{8} ln (ch(4J)),~~~~~~~  \nonumber\\
J_{4}= \frac{1}{8} ln (ch(4J)/ch^{4}(2J)) . \label{J2J4}
\end{eqnarray}
The formulas (\ref{A}) and (\ref{J2J4}) express the solution
through the coupling $J$. It is easy to see that 
\be
A \geq 1,~~~~~~~J_1 = J_{2} \geq 0,~~~~~~~ J_{4} \leq 0.
\ee

Let us rewrite our basic relation in the form
\be
e^{-J_{2} ( \sigma_{1} \sigma_{2} + 
\sigma_{2}\sigma_{3}+\sigma_{3}\sigma_{4} + \sigma_{4}\sigma_{1})}
e^{-J_{2} (\sigma_{1}\sigma_{3} + \sigma_{2}\sigma_{4})}
\frac{1}{2} \sum_{\sigma = \pm 1} e^{J \sigma (\sigma_{1}+ \sigma_{2}+ 
\sigma_{3}+ \sigma_{4} )} = 
A e^{J_{4} (\sigma_{1}\sigma_{2}\sigma_{3}\sigma_{4})} \label{decor}
\ee
where as we have seen $J_{4} \leq 0$. The physical interpretation of 
the last formula is as follows: the initial direct and diagonal antiferromagnetic
interactions $J_1(J) = J_2(J)$ between spins $\sigma_{i}, i=1,2,3,4$ 
together with the interaction $J$ with the central spin $\sigma$ 
generate effective four spin interaction 

\be
e^{J_{4} (\sigma_{1}\sigma_{2}\sigma_{3}\sigma_{4})},~~~~~~~ J_{4} \leq 0.
\ee
Unpleasant feature of this result is that the coupling constant $J_4$ is 
negative while we need it to be positive. To generate four spin 
interaction with positive, ferromagnetic coupling constant $J_4$
we have to change the interaction of one of the spins, let us say 
$\sigma_1 \rightarrow -\sigma_1$ in the formula (\ref{decor})
\be
e^{J_{2} ( \sigma_{1} \sigma_{2} - 
\sigma_{2}\sigma_{3} - \sigma_{3}\sigma_{4} + \sigma_{4}\sigma_{1})}
e^{J_{2} (\sigma_{1}\sigma_{3} - \sigma_{2}\sigma_{4})}
\frac{1}{2} \sum_{\sigma = \pm 1} e^{J \sigma (-\sigma_{1}+ \sigma_{2}+ 
\sigma_{3}+ \sigma_{4} )} = 
A e^{-J_{4} (\sigma_{1}\sigma_{2}\sigma_{3}\sigma_{4})} \label{dec}
\ee
where now the four-spin interaction comes with the right positive 
sign $-J_4 \geq 0$.

\begin{figure}
\centerline{\hbox{\psfig{figure=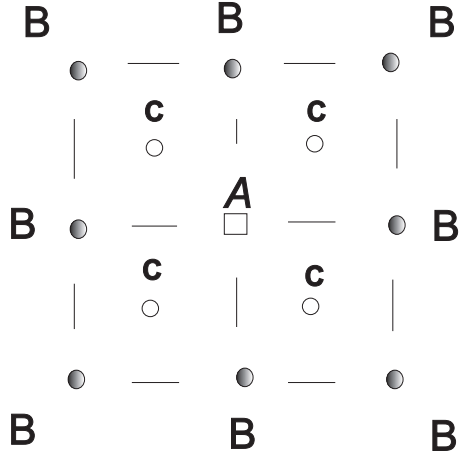,height=6cm,angle=00}}}
\caption[fig7]{Direct interactions AB and  BC are ferromagnetic,
BB and CA are antiferromagnetic. Diagonal interaction AB is 
ferromagnetic and BB is antiferromagnetic. Their intensities are 
given by the formula (\ref{J2J4}). }
\label{fig7}
\end{figure}

The interpretation of the last formula is as follows: one should 
introduce three types of spin-atoms $A,B,C$ as it is shown on Figure 
\ref{fig7} with corresponding interactions between them and   
then after integration over central spin one can see that effective 
interaction is of the type (\ref{fourspin}). This structure can 
be periodically extended to the whole three-dimensional lattice. 
For that one should also use the structure similar to the one shown on 
Figure \ref{fig7}, in which $A$ and $B$ spin-atoms have been interchanged.

\section{Discussion}

As we already discussed in the introduction the  high density magnetic 
recording systems require a storage of information in $nm$ scale,
but fluctuations on $nm$ scale will produce damages which are   
difficult to prevent. We will face this fundamental problem in the near future. 
We have seen that at least theoretically one can construct lattice crystal 
with specially tuned interactions which has exponentially 
degenerate vacuum state and it is suggested that it can 
store information in a form of different vacuum states. 
The planes of flipped spins 
representing different vacuua can in principle be of atomic scale.  

The Ising type spin systems can only mimic real magnetic 
materials and one should think about similar construction 
involving interaction between magnetic domains, but from the other 
side one can also think about materials in which electron or
nuclear spin interaction is organized in a proper way. 

We are facing close phenomena with computer circuits as well.
As the components of the computer circuits become very small, 
in the extreme limit they can approach the atomic scale. Their 
description in the limit of atomic scale should be quantum-mechanical 
\cite{bremerman} and in recent years there have been
intensive studies in the physical limitations of the computational
process \cite{bennett,benioff} which we shall review in Appendix. 

In conclusion I would like to acknowledge Professor E. Paschos for  
discussions and kind hospitality at Dortmund University where part of this
work was completed and to Professor D.Niarchos for pointing out the
problems in $nm$ magnetic recording systems to me \cite{nano}.
This work was supported in part by the EEC Grant no. HPRN-CT-1999-00161
and ESF Network "Geometry and Disorder: From Membranes to Quantum Gravity"
\cite{desy}.

\section{Appendix}

Quantum computers were suggested and analyzed by Benioff and Feynman
\cite{benioff,feynman}. Typically they consist of $N$ interacting 
Ising type two-state spin systems. The initial input state
of quantum computer is a quantum binary string. Computation is accomplished
through the unitary evolution. In the course of computation the intermediate
states are in general superpositions of binary strings.
The theoretical importance of quantum computers comes from the 
realization of the fact that quantum computation can be exponentially
faster than the best known classical algorithms. The most important
examples are quantum algorithms for integer factorization, the 
discrete logarithm and searching in unsorted database \cite{deutsch}.

Although the theory is fairly well understood, the actual building of
quantum computer is extremely difficult. The measurement 
of the state of the quantum computer lead to obstacles in making 
computation reversible \cite{zurek}. The other problem is that 
unknown quantum state cannot be perfectly duplicated. Nevertheless it 
was demonstrated that quantum error correction 
is possible in order to protect quantum information against corruption
\cite{shor}. Quantum teleportation and superdeuce coding 
was developed in \cite{benneff}.   
The problem of maintaining the coherence in the process of quantum 
computation was discussed in \cite{unruh}.

There is an increasing interest in practical realization of 
quantum computers. One of the ideas is to exploit 
quantum effects to build molecular-level computers , that is to induce 
parallel logic in arrays of quantum dots and in molecules
\cite{obermayer}. 
Real physical implementation comes with ions traps: laser cooling 
and thermal isolation of the gaseous Bose-Einstein condensate. The ion 
can be used to operate quantum logic gate that couples the hyperfine
splitting of a single trapped ion $Be_{+}$ to its oscillation
modes in the ion trap \cite{monroe}. 
Optical cavities have been used in the other setup: quantum phase gate, 
was demonstrated for photon pair coupled by a single atom in a quantum 
electrodynamics cavity. The control and target bits of quantum phase gate 
are two photons of different optical frequency, passing together through a 
low-loss QED cavity few microns long\cite{turchette}.

But most promising is probably the bulk nuclear magnetic resonance technique:
nuclear spins act as quantum bits, and are particularly suited to this
role because of their natural isolation from environment 
\cite{gershenfield}.

\vfill


\begin{thebibliography}{99}

\bibitem{domb}C.Domb. Adv.Phys. 9 (1960) 149\\
C.Fan and F.Y.Wu. Phys.Rev. 179 (1969) 560\\
F.Y.Wu. Phys.Rev. 183 (1969) 604\\
M.E.Fisher and W.Selke. Phys.Rev.Lett. 44 (1980) 1502\\
W.Selke. Phys.Rep. 170 (1988) 213\\
D.P.Landay and K.Binder. Phys.Rev. B31 (1985) 5946\\
A.Cappi, P.Colangelo, G.Gonella and A.Maritan. Nucl.Phys.B370 (1992) 659\\
E.N.M.Cirillo and G.Gonella. J.Phys. A28 (1995) 867\\
G.Gonella, S.Lise and A.Maritan. Europhys.Lett. 32 (1995) 735
\bibitem{fischer}K.H.Fischer and J.A.Hertz. Spin Glasses. 
Cambridge Studies in Magnetism. (Cambridge University Press, 1991)\\
Ruderman and Kittel. Phys.Rev. 96 (1954) 99\\
Kasuya. Prog.Theor.Phys. 16 (1956) 45\\
Yosida Phys.Rev. 106 (1957) 893\\
K.Binder Z.Phys. B36 (1979) 161 
\bibitem{bohn}H.G.Bohn and et.al. Phys.Rev. B22 (1980) 5447
\bibitem{sherrington}D.Sherrington and E.P.Anderson. J.Phys.F5 (1975) 965
\bibitem{weg}G.K.Savvidy and F.J.Wegner. Nucl.Phys.B413(1994)605
\bibitem{sav1}G.K.Savvidy and K.G.Savvidy. Phys.Lett. B324 (1994) 72
\bibitem{sav2}G.K. Savvidy and K.G.Savvidy. Phys.Lett. B337 (1994) 333;\\
             G.K.Savvidy, K.G.Savvidy and P.G. Savvidy. Phys.Lett. A221 (1996) 233
\bibitem{weg1}R.Pietig and F.J.Wegner. Nucl.Phys. B466 (1996) 513; B525 (1998) 549 
\bibitem{amb} G.K. Savvidy and K.G. Savvidy. Mod.Phys.Lett. A8 (1993) 2963\\
                R.V. Ambartzumian, G.K. Savvidy , K.G. Savvidy
                and G.S. Sukiasian. Phys. Lett. B275 (1992) 99\\
               G.K. Savvidy and K.G. Savvidy. Int. J. Mod. Phys. A8 (1993) 3993
\bibitem{thordur}T.Jonsson and G.K.Savvidy. Phys.Lett. B449 (1999) 253;
Nucl.Phys. (2000) 
\bibitem{bathas}G.K.Bathas, E.Floratos, G.K.Savvidy and K.G.Savvidy.
Mod.Phys.Lett. A10 (1995) 2695\\
D.Johnson and R.K.P.C.Malmini. Phys.Lett. B378 (1996) 87\\
M.Baig, D.Espriu, D.Johnson and R.K.P.C.Malmini. J.Phys. A30 (1997) 7695\\
E.N.M.Cirillo, G.Gonella, D.Johnson and A.Pelizzola. Phys.Lett. A226 (1997) 59
\bibitem{johnson}M.Baig, D.Espriu, D.Johnson and R.K.P.C.Malmini. 
J.Phys. A30 (1997) 407\\
A.Lipowski and D.Johnson. J.Phys. A30 (1997) 7365; cond-mat/9812098; \\
cond-mat/9910370
\bibitem{felix} F.Ritort. Phys.Rev.Lett. 75 (1995) 1190
\bibitem{wannier}G.H.Wannier. Rev.Mod.Phys.17 (1945) 50\\
                 C.Domb. Adv.Phys. 9 (1960) 149\\
                 I.Syozi. Prog.Theor.Phys. 6 (1951) 306\\
                 S.Naya.  Prog.Theor.Phys. 11 (1954) 53
\bibitem{nano}ESF Programme "Nanomagnetism and growth 
processes on vicinal surfaces (NANOMAG)"
http://www.esf.org/physical/pp/NANOMAG/nanomaga.htm 
\bibitem{desy}ESF Network "Geometry and Disorder: From Membranes to Quantum Gravity" 
http://www.esf.org/physical/pn/Geometry/Geometryb.htm
\bibitem{bremerman}H.J.Bremerman, "Limitations
on Data Processing Arising from Quantum Theory", in Self Organizing
Systems, M.C.Yovits,G.T.Jacobi and G.D.Goldstein, eds. (Spartan Books,
Washington, D.C.,1967)\\
R.Landauer, IBM J.Res.Dev. 5 (1961) 183\\
R.Landauer and J.W.F.Woo. J.Appl.Phys. 42 (1971) 2301\\
R.W.Keyes and R.Landauer. IBM J.Res.Dev. 14 (1970) 152\\
J.D.Bekenstein. Phys.Rev.Lett. 46 (1981) 623
\bibitem{bennett}C.H.Bennett, 
IBM J.Res.Dev. 17 (1973) 525\\
D.Deutsch, Phys.Rev.Lett. 48 (1982) 286\\
R.Landauer, Int.J.Theor.Phys. 21 (1982) 43\\
E.Fredkin and T.Toffili, Int.J.Theor.Phys. 21 (1982) 219\\
P.Benioff, J.Stat.Phys. 22 (1980) 563; 
\bibitem{benioff}P.Benioff, "Quantum mechanical Hamiltonian 
Models of Turing Machines".  J.Stat.Phys. 29 (1982) 515;
Phys.Rev.Lett. "Quantum mechanical models that dissipate 
no energy",  48 (1982) 1581
\bibitem{feynman}R.P.Feynmann. Int.J.Theor.Phys. 21 (1982) 467;
Opt.News 11 (1985) 11; Found.Phys. 16 (1986) 507
\bibitem{deutsch}D.Deutsch, "Quantum theory, the Church-Turing
principle and the universal quantum computer". Proc.R.Soc.Lond. A400 (1985) 97\\
P.Shor, "Algoriphms for quantum computation: discrete logarithms and factoring".
Proc.35-th.Annu.Symp. on Foun. of Computer Science 124-134.(IEEE Comp.Soc.Press.
Los Alamitos, CA,1994)\\
D.Deutsch and R.Jozsa, "Rapid solution of problems by quantum computation".
Proc.R.Soc.London. A439 (1992) 553\\
L.K.Grover, "Quantum computers can search arbitrary large date bases
by a single query". Phys.Rev.Lett. 79 (1997) 4709
\bibitem{zurek}W.H.Zurek.Phys.Rev.Lett. 53 (1984) 391; Phys.Rev.D26 (1981) 1862\\
W.K.Wooters and W.H.Zurek. Nature 299 (1982) 802
\bibitem{shor}P.Shore. Phys.Rev A52 (1995) 2493\\
A.M.Steane. Phys.Rev.Lett. 77 (1996) 793;Proc.R.Soc.Lond. A452 (1996) 1551 
\bibitem{benneff} C.H.Benneff and et.al. Phys.Rev.Lett. 70 (1993) 1895\\
C.H.Benneff and S.J.Wiesner.Phys.Rev.Lett. 69 (1992) 2881
\bibitem{unruh}R.Landauer.Philos.Trans.R.Soc.London 353 (1995) 367\\
W.G.Unruh. Phys.Rev. A51 (1995) 992
\bibitem{obermayer}K.Obermayer, W.G.Teich and G.Mahler. Phys.Rev B37 (1988) 8096;
Phys.Rev B37 (1988) 8111; Phys.Rev A45 (1992) 3300\\
S.Lloyd. Science 261 (1993) 1569\\
F.L.Carter. Molecular Electronic Devices I,II. (Dekker, New York, 1982, 1987)\\
D.K.Ferry, J.R.Barker and C.Jacobini. Granular Nanoelectronics,
NATO Advanced Study Institute on Physics of Granular 
Nanoelectronics. (Plenum Press, New York, 1991)\\
K.E.Drexler. Nanosystems: molecular machinery, manifacturing and 
computation (Wiley, New York, 1992) \\
G.Toulouse. Biology and computation: a physicist choice.
(World Scientific, Singapore, 1994)
\bibitem{monroe}C.Monroe and et.al. Phys.Rev.Lett. 75 (1995) 4714\\
J.I.Cirac and P.Zoller. Phys.Rev.Lett. 74 (1995) 4091
\bibitem{turchette}Q.Turchette and et.al. Phys.Rev.Lett. 75 (1995) 4710
\bibitem{gershenfield}N.Gershenfield and I.L.Chuang. Science 275 (1997) 350\\
D.G.Cory,A.F.Fahning and T.F.Havel. Proc.Nat.Acad.Sci.USA 94 (1997) 1634\\
I.L.Chuang and et.al. Nature 393 (1998) 143\\
J.A.Jones,M.Mosca and R.H.Hansen. Nature 393 (1998) 344\\
D.J.Wineland. Phys.Rev.Lett. 75 (1995) 4714\\
I.L.Chuang, N.Gerschenfield and M.Kubinec. Phys.Rev.Lett. 80 (1998) 3408\\
D.G.Cory and et.al. Phys.Rev.Lett. 81 (1998) 2152; Physica 120D (1998) 82\\
R.Laflamme and et.al. Proc.Trans.R.Soc.London. A356 (1998) 1941\\
N.Gisin and S.Popescu. Phys.Rev.Lett. 83 (1999) 432\\
A. Lorke and et.al. Phys. Rev. Lett. 84 (2000) 2223 \\
P. G. Savvidis,J.J.Baumberg and et.al. Phys. Rev. Lett. 84 (2000) 1547 

\end{thebibliography}
\end{document}